\documentclass[twocolumn,showpacs,superscriptaddress,aps]{revtex4}
\usepackage{amsmath,graphicx,epsfig}

\begin{document}

\title{Detection of a single cobalt microparticle with a microfabricated atomic magnetometer}

\author{D. Maser}\email{dmaser@berkeley.edu}
\affiliation{Department of Physics, University of California at Berkeley, Berkeley, California 94720-7300}
\author{S. Pandey}
\affiliation{Department of Physics, University of California at Berkeley, Berkeley, California 94720-7300}
\author{H. Ring}
\affiliation{Department of Chemistry, University of California at Berkeley, Berkeley, California 94720-3220}
\author{M.\ P.\ Ledbetter}
\affiliation{Department of Physics, University of California at Berkeley, Berkeley, California 94720-7300}
\author{S.\ Knappe}
\affiliation{National Institute of Standards and Technology, Time and Frequency Division, 325 Broadway, Boulder, Colorado 80305}
\affiliation{Department of Physics, University of Colorado at Boulder, Boulder, Colorado 80309}
\author{J.\ Kitching}
\affiliation{National Institute of Standards and Technology, Time and Frequency Division, 325 Broadway, Boulder, Colorado 80305}
\author{D.\ Budker}
\affiliation{Department of Physics, University of California at Berkeley, Berkeley, California 94720-7300}
\affiliation{Nuclear Science Division, Lawrence Berkeley National Laboratory, Berkeley, California 94720}

\date{\today}

\begin{abstract}
We present magnetic detection of a single, $\rm 2~ \mu m$ diameter cobalt microparticle using an atomic magnetometer based on a microfabricated vapor cell. These results represent an improvement by a factor of $10^5$ in terms of the detected magnetic moment over previous work using atomic magnetometers to detect magnetic microparticles.  The improved sensitivity is due largely to the use of small vapor cells.  In an optimized setup, we predict detection limits of $\rm 0.17~ \mu m^3$.
\end{abstract}

\pacs{07.55.Jg, 75.50.Tt, 85.75.Ss}

\maketitle

\section{Introduction}

Magnetic microparticles and methods for their detection show promise in a variety of fields, most notably in biophysical and medical applications \cite{Gijs,Pamme,Pankhurst}. Such particles can be coated with biomolecules, enabling magnetic labeling, manipulation, and separation of cells. For example, proteins in solution can be detected at attomolar concentrations by binding magnetic microparticles to antibodies through the use of bio-bar-codes \cite{Nam}. That work employed scanometric DNA detection which binds DNA and gold nanoparticles to magnetic microparticles and measures light scattered from the gold nanoparticles, rather than directly detecting the magnetic moment of the magnetic microparticles.

This work follows the work of Refs. \cite{Xu,Xu2,Yao,Yao2}, using a sensitive atomic magnetometer based on an alkali vapor cell to directly detect the magnetic moment of magnetic microparticles. These prior demonstrations employed blown glass cells with dimensions of about 1 cm, and sensitivities of about ${\rm  1~ nG/\sqrt{Hz}}$ with a bandwidth on the order of $\rm 10~ Hz$. Single ferromagnetic cobalt particles with a diameter of $150~{\rm \mu m}$ were detected, and based on measured signal-to-noise ratio, the authors projected detection limits as small as $20~{\rm \mu m}$. Recent developments in microfabrication technology and the use of the spin-exchange relaxation-free (SERF) \cite{Kominis2003} regime have yielded sensors a factor of $10^2$ to $10^3$ smaller in volume with demonstrated sensitivities better than ${\rm  0.1~ nG/\sqrt{Hz}}$ and bandwidth in excess of $\rm 300~ Hz$ \cite{Shah,Ledbetter}.

Here, employing such a microfabricated SERF magnetometer, we detect the magnetic signal produced by a single cobalt microparticle of less than $\rm 2~ \mu m$ in diameter with a signal-to-noise ratio of roughly 10.  In terms of magnetic moment, this represents an improvement by more than a factor of $10^5$ compared to the work reported in Ref. \cite{Xu}. The improvement in sensitivity is primarily due to more closely matching the volume of vapor cell and microparticle without losing magnetometric sensitivity.   Newly emerging techniques in magnetometry based on either nitrogen-vacancy (NV) centers in diamond \cite{Taylor} or a spinor Bose-Einstein condensate (BEC) \cite{Higbie} have yielded sensors with much smaller volumes, with dimensions on the order of nanometers and $10~{\rm \mu m}$, respectively, potentially offering dramatically improved sensitivity to nanometer scale magnetic particles.

\section{Experimental setup and procedure}

\begin{figure}[tb]
	\includegraphics{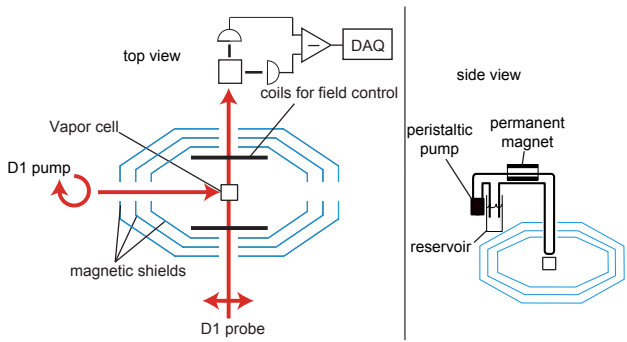}\\
	\caption{Experimental setup using suspension of particles in a fluid. A ${\rm ^{87}Rb}$ vapor cell is housed inside a set of magnetic shields. The vapor cell is optically pumped via circularly polarized light resonant with the D1 transition and probed via linearly polarized light, tuned about 100 GHz off-resonance with the D1 transition, yielding sensitivity to magnetic fields transverse to both beams. A peristaltic pump transports a solution of fluid carrying the magnetic particle through teflon tubing adjacent to the vapor cell. In an alternate setup (not shown), we used a handheld syringe to move the particle back and forth past the magnetometer. Prior to acquiring data, the particle was magnetized in a $6$ kG magnet.}
	\label{Fig:slurry}
\end{figure}

The experimental setup for detecting magnetic microparticles entrained in a liquid is shown in Fig. \ref{Fig:slurry}.
The magnetometer operates in the spin-exchange relaxation-free (SERF) regime. The central component of the magnetometer is a microfabricated vapor cell with dimensions of ${\rm 1~mm \times 2~mm\times 3~mm}$, containing ${\rm ^{87}Rb}$ and 1300 torr of ${\rm N_2}$ buffer gas.  The alkali vapor is optically pumped with a circularly polarized laser beam tuned to the center of the pressure-broadened D1 transition.  Alkali spin-precession is probed with a linearly polarized laser beam tuned about 100 GHz to the blue of the center of the pressure broadened D1 transition.  A magnetic field orthogonal to both beams rotates the alkali spin polarization into the direction of the probe beam.  The optical rotation of the probe beam is proportional to the component of spin-polarization along the probe beam, and thus proportional to the magnetic field over a range of about ${\rm \pm 400~ \mu G}$. Data were acquired with a sampling rate of 200 Hz, with RMS noise of about 6.8 nG, yielding a sensitivity of about ${\rm 0.5~ nG/\sqrt{Hz}}$. Signals were acquired by entraining the particle in water that flowed past the magnetometer in 3.1 mm OD teflon tubing. Flow was accomplished via either a peristaltic pump for continuous flow, or a handheld syringe for oscillating flow.

\section{Data for cobalt microparticles}

\begin{figure}[tb]
	\includegraphics{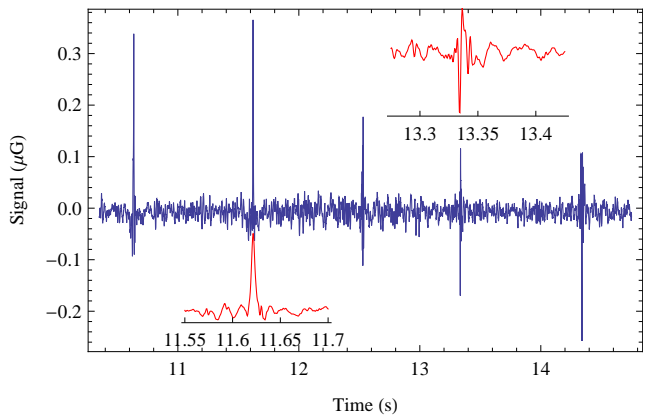}\\
	\caption{Signal from a single cobalt microparticle (diameter $\approx{\rm 2~\mu m}$), repeatedly drawn across the detection range of the vapor cell using a syringe, with two insets, each showing the lineshape of one particular spike.}
	\label{Fig:Cobaltsignals}
\end{figure}

\begin{figure}[tb]
  \includegraphics{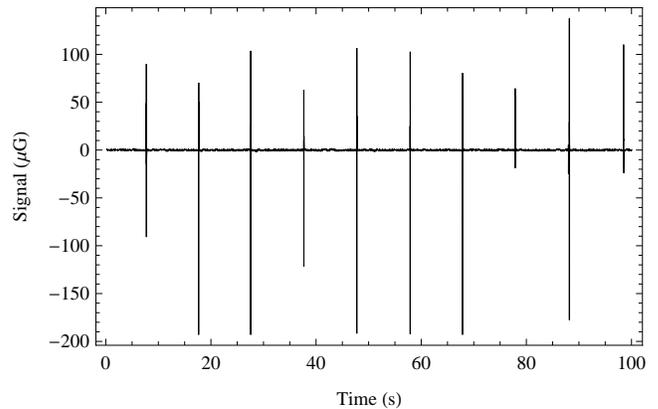}\\
  \caption{Signal from a single cobalt particle (diameter $\approx{\rm 20~\mu m}$), encapsulated in an epoxy pellet, circulating through the system via a peristaltic pump.}
  \label{Fig:Cobaltsignals_old}
\end{figure}

Signals from a single cobalt particle using the syringe method are shown in Fig. \ref{Fig:Cobaltsignals}. To acquire these data, a droplet of water containing roughly 20 particles (determined using an optical microscope) was added into a test tube with water, forming a dilute ``slurry''.  The test tube was inserted into a $6$ kG permanent magnet to magnetize the cobalt particles.  The slurry was then sucked into the teflon tube and past the magnetometer.  Once a spike in the magnetic field was observed (as shown in Fig. \ref{Fig:Cobaltsignals}), the particle was made to oscillate past the vapor cell using the syringe so that a single particle's magnetic field could be repeatedly measured. One may question whether the observed signal is really due to a single, ${\rm 2~\mu m}$ particle, as it is conceivable that the particles could clump together during the magnetization procedure. To ensure that the particles did not clump together, we performed an auxiliary test, where we placed the droplet on a microscope slide, inserted it into the magnet, and examined it under an optical microscope. We observed no clumping.

Magnetic-field measurements as the particle oscillates past the vapor cell are shown in Fig. \ref{Fig:Cobaltsignals}. Information regarding the orientation of the particle can be extracted from the time variation of the transient in the inset \cite{Yao2}. The spikes were typically $\rm 0.2~ \mu G$ to $\rm 0.4~\mu G$ in magnitude, whereas the RMS noise never exceeded $\rm 0.02~ \mu G$. For comparison, data from a ${\rm 20~ \mu m}$ particle (roughly 1000 times larger in volume) are shown in Fig. \ref{Fig:Cobaltsignals_old}. These data were acquired using the peristaltic pump, hence the regular spacing between the peaks.

The observed signals are consistent with what one would expect, assuming full magnetization. The maximum magnetic field at the location of the sensor due to a microparticle is (in Gaussian units)
\begin{equation}
B=\dfrac{2m}{r^3}=\dfrac{1.72 \mu_B N_A \rho V}{\mathcal{M} r^3}=0.30-0.82~ \mu \mathrm{G},
\label{eq}
\end{equation}
where $m$ is the magnetic moment of the particle, $1.72 \mu_B$ is the magnitude of the magnetic moment per atom of cobalt in Bohr magnetons, $\rho$ is the mass density of cobalt, $V$ is the volume of the particle, $\mathcal{M}$ is the molar mass, $N_A$ is Avogadro's number, and $r$ indicates the approximate distance from the center of the vapor cell to the Teflon tube. The range of values given on the right hand side of Eq. \ref{eq} corresponds to $r$ ranging from 2.5 mm to 3.5 mm. The peak-to-peak size of the spikes seen in Fig. \ref{Fig:Cobaltsignals} is roughly half of the maximum magnetic field produced by the particle, consistent with the estimate given by Eq. \ref{eq}.

Using the same equation, we can estimate the detection limit, assuming that the smallest discernable signal should be roughly twice the magnitude of the RMS noise ($\rm 0.04~ \mu G$). The corresponding limit is found to be $\rm 0.17~ \mu m^3$. This is based on the sensitivity of the current setup; sensitivities roughly 10 times better have been achieved in similarly sized devices, indicating further room for improvement.

\section{Conclusions}
We have shown that ferromagnetic particles with diameters of roughly $\rm 2~\mu m$ can be detected with an atomic magnetometer. The sensitivity of this magnetometer is better by a factor of approximately two compared to that demonstrated in Ref. \cite{Xu}; additionally, the smaller size of the sensor and higher bandwidth improve the sensitivity further. This allows the measurement of signals that are a factor of 15 smaller in terms of the magnetic field size, and a factor of $10^5$ smaller in terms of the volume and magnetic moment of the ferromagnetic particles.

Magnetic microparticles in the micron and nanometer scale are used in drug targeting, cell separation, MRI, radiotherapy, drug delivery, hyperthermic treatment, magnetic separation of bacteria, viruses, and parasites, and blood detoxification \cite{schutt}. Hence the ability to detect magnetic microparticles demonstrated here may find direct application in some of these biomedical techniques as well as other applications in industry (for example, for process control), and in other basic scientific research. For example, sea slugs are thought to navigate via ferromagnetic microparticles.  The technique demonstrated here may enable direct detection of the magnetically sensitive tissues in such organisms.

\section{Acknowledgements}
Research was supported by the US Department of Energy, Office of Basic Energy Sciences, Division of Materials Sciences and Engineering under Contract \#DE-AC02-05CH11231 (H.R.), by the National Science Foundation under award \#CHE-0957655 (D.M., D.B., and M.P.L.), and by the National Institute of Standards and Technology (S.K. and J.K.). We gratefully acknowledge discussions with S. Xu and L. S. Bouchard. This work is a partial contribution of NIST, an agency of the US government, and is not subject to copyright.


\begin{thebibliography}{10}
\bibliographystyle{aipauth4-1}

\bibitem{Gijs}
M.~A.~M. Gijs.
\newblock {Magnetic bead handling on-chip: new opportunities for analytical
  applications}.
\newblock {\em {Microfluidics and Nanofluidics}}, {1}({1}):{22--40}, {Nov}
  {2004}.

\bibitem{Kominis2003}
I.~K. Kominis, T.~W. Kornack, J.~C. Allred, and M.~V. Romalis.
\newblock {A subfemtotesla multichannel atomic magnetometer}.
\newblock {\em {Nature}}, {422}({6932}):{596--599}, {Apr 10} {2003}.

\bibitem{Ledbetter}
M.~P. Ledbetter, I.~M. Savukov, D.~Budker, V.~Shah, S.~Knappe, J.~Kitching,
  D.~J. Michalak, S.~Xu, and A.~Pines.
\newblock {Zero-field remote detection of NMR with a microfabricated atomic
  magnetometer}.
\newblock {\em {Proceedings of the National Academy of Sciences of the United
  States of America}}, {105}({7}):{2286--2290}, {Feb 19} {2008}.

\bibitem{Nam}
J.~M. Nam, C.~S. Thaxton, and C.~A. Mirkin.
\newblock {Nanoparticle-based bio-bar codes for the ultrasensitive detection of
  proteins}.
\newblock {\em {Science}}, {301}({5641}):{1884--1886}, {Sep 26} {2003}.

\bibitem{Pamme}
N.~Pamme.
\newblock {Magnetism and microfluidics}.
\newblock {\em {Lab on a Chip}}, {6}({1}):{24--38}, {2006}.

\bibitem{Pankhurst}
Q.~A. Pankhurst, N.~K.~T. Thanh, S.~K. Jones, and J.~Dobson.
\newblock {Progress in applications of magnetic nanoparticles in biomedicine}.
\newblock {\em {Journal of Physics D-Applied Physics}}, {42}({22}), {Nov 21}
  {2009}.

\bibitem{schutt}
W.~Schutt, C.~Gruttner, U.~Hafeli, M.~Zborowski, J.~Teller, H.~Putzar, and
  C.~Schumichen.
\newblock {Applications of magnetic targeting in diagnosis and therapy -
  Possibilities and limitations: A mini-review}.
\newblock {\em {Hybridoma}}, {16}({1}):{109--117}, {Feb} {1997}.

\bibitem{Shah}
Vishal Shah, Svenja Knappe, Peter D.~D. Schwindt, and John Kitching.
\newblock {Subpicotesla atomic magnetometry with a microfabricated vapour
  cell}.
\newblock {\em {Nature Photonics}}, {1}({11}):{649--652}, {Nov} {2007}.

\bibitem{Taylor}
J.~M. Taylor, P.~Cappellaro, L.~Childress, L.~Jiang, D.~Budker, P.~R. Hemmer,
  A.~Yacoby, R.~Walsworth, and M.~D. Lukin.
\newblock {High-sensitivity diamond magnetometer with nanoscale resolution}.
\newblock {\em {Nature Physics}}, {4}({10}):{810--816}, {OCT} {2008}.

\bibitem{Higbie}
M.~Vengalattore, J.~M. Higbie, S.~R. Leslie, J.~Guzman, L.~E. Sadler, and D.~M.
  Stamper-Kurn.
\newblock {High-resolution magnetometry with a spinor Bose-Einstein
  condensate}.
\newblock {\em {Physical Review Letters}}, {98}({20}), {May 18} {2007}.

\bibitem{Xu}
S.~Xu, M.~H. Donaldson, A.~Pines, S.~M. Rochester, D.~Budker, and V.~V.
  Yashchuk.
\newblock {Application of atomic magnetometry in magnetic particle detection}.
\newblock {\em {Applied Physics Letters}}, {89}({22}), {Nov 27} {2006}.

\bibitem{Yao2}
Li~Yao, Andrew~C. Jamison, and Shoujun Xu.
\newblock {Scanning Imaging of Magnetic Nanoparticles for Quantitative
  Molecular Imaging}.
\newblock {\em Angewandte Chemie International Edition}, 49(41):7493--7496,
  2010.

\bibitem{Yao}
Li~Yao and Shoujun Xu.
\newblock {Long-Range, High-Resolution Magnetic Imaging of Nanoparticles}.
\newblock {\em {Angewandte Chemie International Edition}},
  {48}({31}):{5679--5682}, {2009}.

\bibitem{Xu2}
Li~Yao and Shoujun Xu.
\newblock {Force-Induced Remnant Magnetization Spectroscopy for Specific
  Magnetic Imaging of Molecules}.
\newblock {\em Angewandte Chemie International Edition}, 50(18):n/a--n/a, 2011.

\end{thebibliography}
\end{document}